\documentclass{anabs}
\usepackage{graphicx}
\usepackage{times}
\Volume{1-2}             
\Year{2003}              
\Month{02}               
\Pagespan{000}{000}      

\begin{document}
\lhead[\thepage]{A.N. Kappes, M., Kerp, J.: A window to the Galactic X-ray halo: The ISM towards the Lockman hole}
\rhead[Astron. Nachr./AN~{\bf 324} (2003) 1/2]{\thepage}
\headnote{Astron. Nachr./AN {\bf 324} (2003) 1/2, 000--000}

\title{A window to the Galactic X-ray halo: The ISM towards the Lockman hole}

\author{M. Kappes, J. Kerp}
\institute{Radioastronomisches Institut der Universit\"at Bonn, Auf dem H\"ugel 71, 53121 Bonn, Germany}

\correspondence{mkappes@astro.uni-bonn.de}

\maketitle

\section{Radiative transfer model for soft X-rays}
The diffuse soft X-ray emission ($E<1$\,keV) of the Milky Way is a sensitive tool to study the distribution
of the photoelectric absorbing ISM. Our field of interest encloses the Lockman hole. It represents the
absolute minimum of $N_\ion{H}{i}$ on the whole sky, accordingly it is considered as the ``window to the
distant universe''. To test this hypothesis we correlate the ROSAT all-sky survey energy bands {\bf R1},
{\bf R2}, {\bf C} and {\bf M} (Snowden et al. 1994) with the Leiden/Dwingeloo \ion{H}{i} 21\,cm--line survey
(Hartmann \& Burton 1997). We choose a large portion of the sky ($60^{\circ} \times 60^{\circ}$) to model
the ROSAT data across a high dynamic range in X--ray intensity and $N_\ion{H}{i}$. Deviations in plasma
emissivity ($\propto n_{\rm e}^2$) or absorption ($N_\ion{H}{i}$) can be identified in the difference map between
model and observational X--ray intensity distribution. Following Kerp et al. (1999) we solve the X-ray
radiative transfer equation:

\begin{equation}
\label{eqn_radiation}
I = I_{\rm LHB} + I_{\rm HALO} \cdot {\rm e}^{- \sigma_\mathrm{h} \cdot N_{\ion{H}{i},h}} + I_{\rm EXTRA} \cdot {\rm e}^{- \sigma_\mathrm{e} \cdot N_{\ion{H}{i},e}}
\end{equation}

Here, $I_{\rm LHB}$ is the X--ray intensity of the Local Hot Bubble, $I_{\rm HALO}$ the diffuse X--ray emission
of the Milky Way halo, and $I_{\rm EXTRA}$ the extragalactic X--ray background. The amount of the absorbing
material is traced by the \ion{H}{i} column density distribution ($N_\ion{H}{i}(l,b)$), $\sigma$ denotes the
photoelectric absorption cross section. Equation \ref{eqn_radiation} has four free ($I_{\rm LHB}; I_{\rm HALO};
T_{\rm LHB}; T_{\rm HALO}$) and two fixed parameters ($I_{\rm EXTRA}$, Barber et al. 1996; $\Gamma$, Hasinger et
al. 2001). To determine the four free parameters we use a fitprocedure (Kappes, Pradas \& Kerp 2002) which
approximates the C, M, C/M, and R1/R2 vs. $N_\ion{H}{i}$ diagrams {\em simultaneously}. This approach allows
us to determine the plasma temperatures and $n_{\rm e}^2$ as a function of $N_\ion{H}{i}$ very accurately.

\begin{figure}
{\includegraphics[scale=0.235]{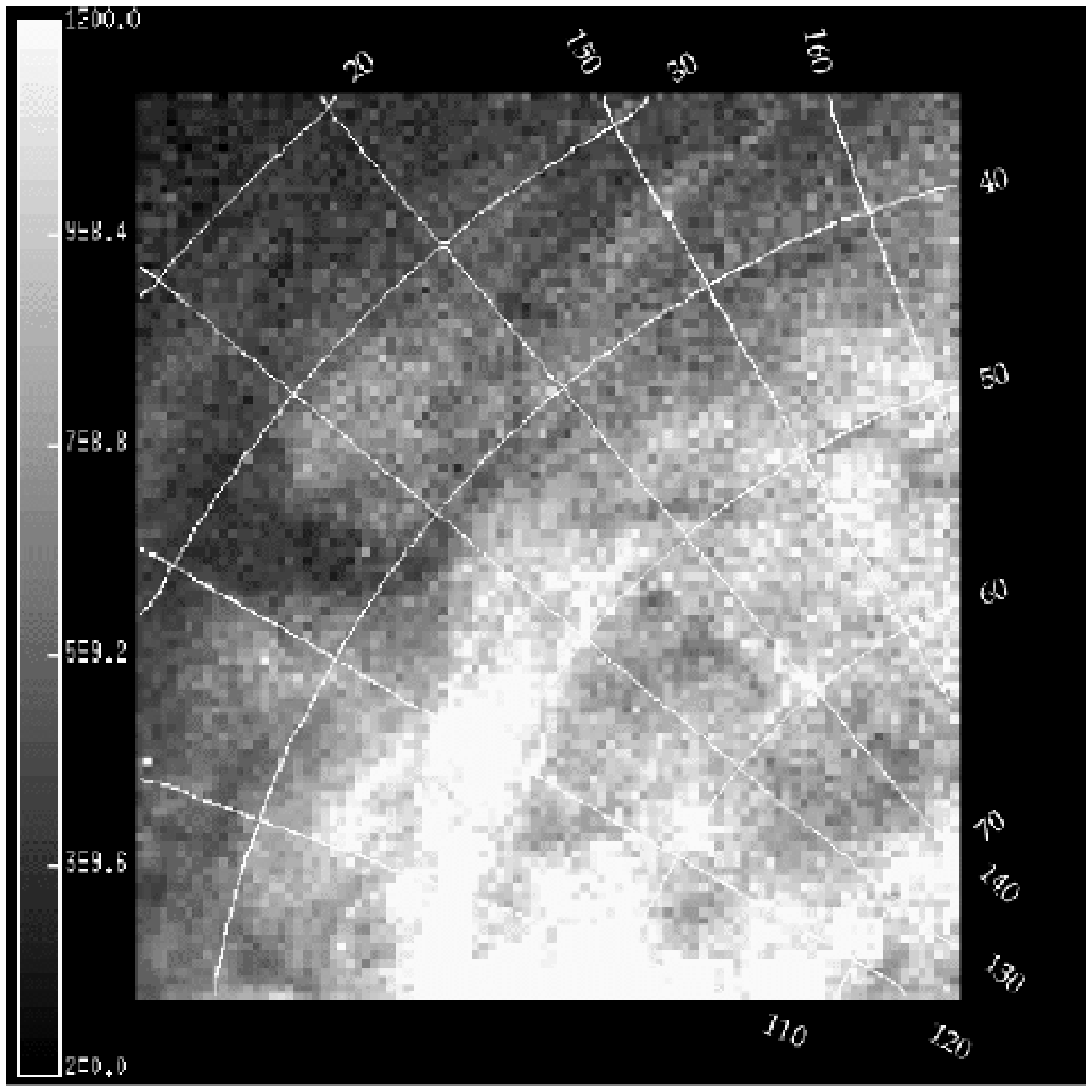}}
{\includegraphics[scale=0.25]{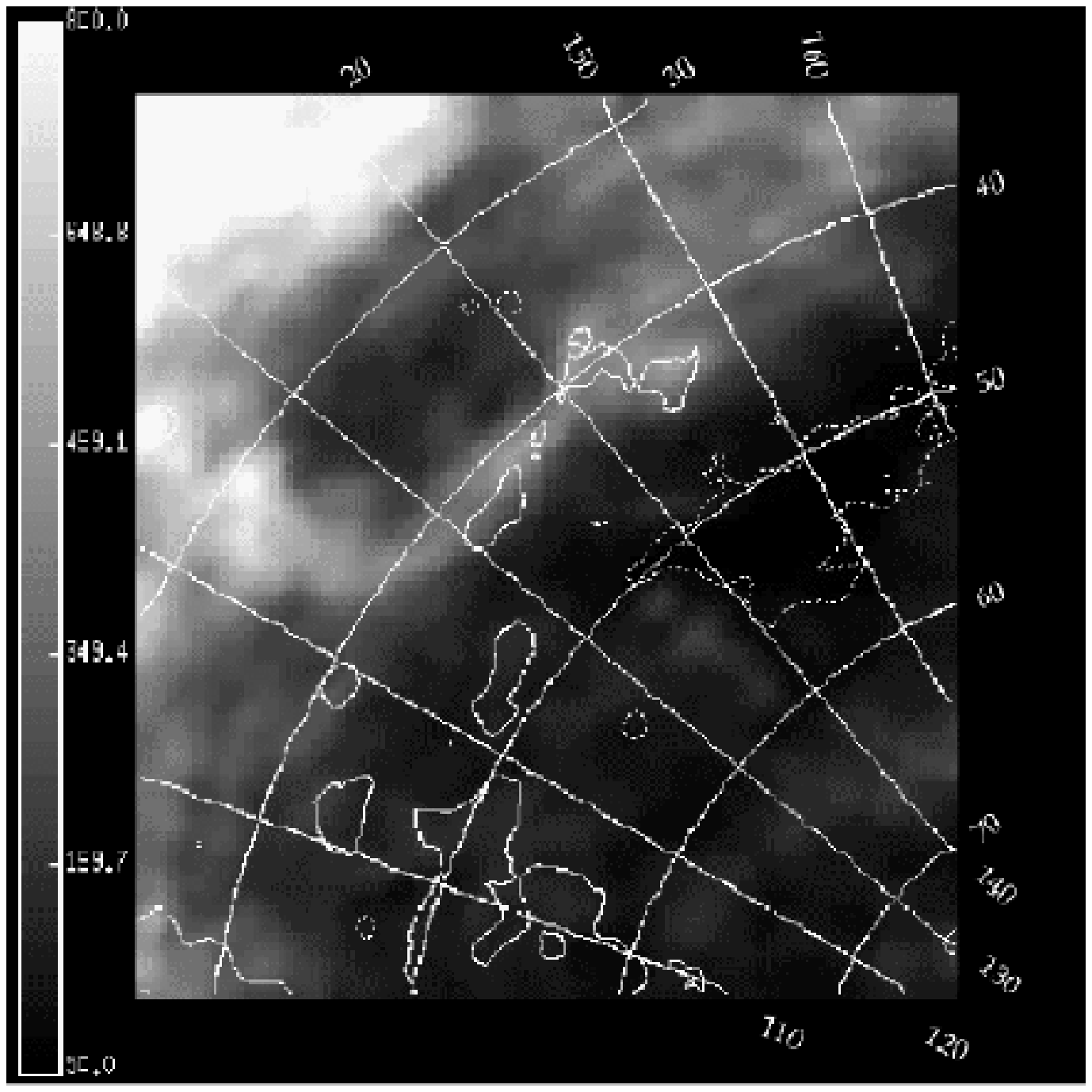}}
\caption{{\bf left:} The observed ROSAT C-band intensity of the selected field.
         {\bf right:} The \ion{H}{i} distribution towards the field of interest. 
         Solid contours encircle regions where the modeled X--ray intensity is 
	 too faint, dashed contours mark too bright regions ($2.5\sigma$). Only
	 the Lockmann hole region is enclosed by the dashed lines, indicating
	 that the $N_\ion{H}{i}$ does not trace the whole ISM.}
\label{xrhi}
\end{figure}

\section{Results}
One of the main results is that $T_{\rm HALO} = 10^{6.2}\,$K$\,> T_{\rm LHB} = 10^{6.0}\,$K. We find
a quantitative agreement between the expected photoelectric absorption and the observed X--ray 
intensity distribution across tens of degrees (see areas encircled by the contour lines in Fig. \ref{xrhi},
right). However, towards the Lockman hole our model fails to reproduce the X--ray intensity distribution.
It appears that the Lockman hole is not as transparent as the \ion{H}{i} data suggest. We propose that
about half of the X--ray absorbing ISM towards the Lockman hole is in form of {\em ionized} hydrogen
rather than \ion{H}{i}. In a forthcoming paper (Kappes et al. in prep) more details will be presented.

\begin{acknowledgements}
  The authors like to thank the Deutsches Zentrum f\"ur Luft- und
  Raumfahrt for financial support under grant No. 50 OR 0103.
\end{acknowledgements}

\end{document}